\documentclass[prx,longbibliography, twocolumn,lengthcheck,superscriptaddress,nopreprintnumbers, nofootinbib]{revtex4-1}

\usepackage{natbib,units}
\usepackage{amsmath,amssymb}
\usepackage{rotating}
\usepackage[usenames]{color}
\usepackage{color}
\usepackage[usenames,dvipsnames]{xcolor}

\usepackage{ulem}
\usepackage[plainpages=false, colorlinks=true, anchorcolor=blue, linkcolor=blue, citecolor=blue, bookmarks=false]{hyperref}
\newcommand{\gw}{{\rm gw}}
\newcommand{\cmb}{{\rm CMB}}
\newcommand{\eq}{{\rm eq}}
\newcommand{\yr}{{\rm yr}}

\newcommand{\jcap}{JCAP}

\def\be{\begin{equation}}
\def\ee{\end{equation}}

\def\bea{\begin{eqnarray}}
\def\eea{\end{eqnarray}}

\begin{document}
\title{Gravitational-wave cosmology across 29 decades in frequency}

\author{Paul~D.~Lasky}
\affiliation{Monash Centre for Astrophysics, School of Physics and Astronomy, Monash University, VIC 3800, Australia}
\email{paul.lasky@monash.edu}

\author{ Chiara~M.~F. Mingarelli}
\affiliation{TAPIR Group, MC 350-17, California Institute of Technology, Pasadena, California 91125, USA}
\affiliation{Max Planck Institute for Radio Astronomy, Auf dem H\"{u}gel 69, D-53121 Bonn, Germany }

\author{Tristan~L. Smith}
\affiliation{Department of Physics and Astronomy, Swarthmore College, Swarthmore, PA 19081, USA}

\author{John T. Giblin, Jr.}
\affiliation{Department of Physics, Kenyon College, Gambier, Ohio 43022, USA}
\affiliation{Department of Physics \& CERCA, Case Western Reserve University, Cleveland, Ohio 44106, USA}

\author{Eric Thrane}
\author{Daniel J. Reardon}
\affiliation{Monash Centre for Astrophysics, School of Physics and Astronomy, Monash University, VIC 3800, Australia}

\author{Robert Caldwell}
\affiliation{Department of Physics \& Astronomy, Dartmouth College, Hanover, NH 03755 USA}

\author{Matthew Bailes}
\affiliation{Centre for Astrophysics and Supercomputing, Swinburne University of Technology, P.O. Box 218, Hawthorn, Victoria 3122, Australia}

\author{N. D. Ramesh Bhat}
\affiliation{International Centre for Radio Astronomy Research, Curtin University, Bentley, Western Australia 6102, Australia}

\author{Sarah Burke-Spolaor}
\affiliation{National Radio Astronomy Observatory Array Operations Center, P.O. Box O, Soccoro, NM 87801, USA }


\author{Shi Dai}
\affiliation{Australia Telescope National Facility, CSIRO Astronomy \& Space Science, P.O. Box 76, Epping, NSW 1710, Australia}
\affiliation{School of Physics and State Key Laboratory of Nuclear Physics and Technology, Peking University, Beijing 100871, China}

\author{James Dempsey}
\affiliation{CSIRO Information Management \& Technology, PO Box 225, Dickson ACT 2602 }

\author{George Hobbs}
\author{Matthew Kerr}
\affiliation{Australia Telescope National Facility, CSIRO Astronomy \& Space Science, P.O. Box 76, Epping, NSW 1710, Australia}

\author{Yuri Levin}
\affiliation{Monash Centre for Astrophysics, School of Physics and Astronomy, Monash University, VIC 3800, Australia}

\author{Richard N. Manchester}
\affiliation{Australia Telescope National Facility, CSIRO Astronomy \& Space Science, P.O. Box 76, Epping, NSW 1710, Australia}

\author{Stefan Os{\l}owski}
\affiliation{Fakult\"{a}t f\"{u}r Physik, Universit\"{a}t Bielefeld, Postfach 100131, 33501 Bielefeld, Germany}
\affiliation{Max Planck Institute for Radio Astronomy, Auf dem H\"{u}gel 69, D-53121 Bonn, Germany }

\author{Vikram Ravi}
\affiliation{Cahill Center for Astronomy and Astrophysics, MC 249-17, California Institute of Technology, Pasadena, CA 91125, USA}

\author{Pablo A. Rosado}
\affiliation{Centre for Astrophysics and Supercomputing, Swinburne University of Technology, P.O. Box 218, Hawthorn, Victoria 3122, Australia}

\author{Ryan M. Shannon}
\affiliation{Australia Telescope National Facility, CSIRO Astronomy \& Space Science, P.O. Box 76, Epping, NSW 1710, Australia}
\affiliation{International Centre for Radio Astronomy Research, Curtin University, Bentley, Western Australia 6102, Australia}

\author{Ren\'ee Spiewak}
\affiliation{Department of Physics, University of Wisconsin-Milwaukee, P.O. Box 413, Milwaukee, WI 53201, USA}

\author{Willem van Straten}
\affiliation{Centre for Astrophysics and Supercomputing, Swinburne University of Technology, P.O. Box 218, Hawthorn, Victoria 3122, Australia}

\author{Lawrence Toomey}
\affiliation{Australia Telescope National Facility, CSIRO Astronomy \& Space Science, P.O. Box 76, Epping, NSW 1710, Australia}

\author{Jingbo Wang}
\affiliation{Xinjiang Astronomical Observatory, Chinese Academy of Sciences, 150 Science 1-Street, Urumqi, Xinjiang 830011, China}

\author{Linqing Wen}
\affiliation{School of Physics, University of Western Australia, Crawley, WA 6009, Australia}

\author{Xiaopeng You}
\affiliation{School of Physical Science and Technology, Southwest University, Chongqing, 400715, China}

\author{Xingjiang Zhu}
\affiliation{School of Physics, University of Western Australia, Crawley, WA 6009, Australia}

\begin{abstract}
Quantum fluctuations of the gravitational field in the early Universe, amplified by inflation, produce a primordial gravitational-wave background across a broad frequency band.  We derive constraints on the spectrum of this gravitational radiation, and hence on theories of the early Universe, by combining experiments that cover 29 orders of magnitude in frequency.  These include Planck observations of cosmic microwave background temperature and polarization power spectra and lensing, together with baryon acoustic oscillations and big bang nucleosynthesis measurements, as well as new pulsar timing array and ground-based interferometer limits.  While individual experiments constrain the gravitational-wave energy density in specific frequency bands, the combination of experiments allows us to constrain cosmological parameters, including the inflationary spectral index, $n_t$, and the tensor-to-scalar ratio, $r$. Results from individual experiments include the most stringent nanohertz limit of the primordial background to date from the Parkes Pulsar Timing Array, $\Omega_\gw(f)<2.3\times10^{-10}$.  Observations of the cosmic microwave background alone limit the gravitational-wave spectral index at 95\% confidence to $n_t\lesssim5$ for a tensor-to-scalar ratio of $r = 0.11$. However, the combination of all the above experiments limits $n_t<0.36$.  Future Advanced LIGO observations are expected to further constrain $n_t<0.34$ by 2020.  When cosmic microwave background experiments detect a non-zero $r$, our results will imply even more stringent constraints on $n_t$ and hence theories of the early Universe.
\end{abstract}

\maketitle
Gravitational-wave astronomy is now a reality.  The LIGO Scientific Collaboration has recently announced the first direct detection of gravitational waves coming from the merger of a binary black hole~\cite{abbott16_detection}.  Other experiments worldwide are ready to measure gravitational radiation across a wide range of frequencies.
From the cosmic microwave background (CMB) to ground-based GW interferometers, these experiments cover more than 21 orders of magnitude in frequency---29 with complementary but {\it indirect} bounds from big bang nucleosynthesis (BBN), CMB temperature and polarization power spectra and lensing, and baryon acoustic oscillations (BAO) measurements.  Each of these experiments is sensitive to a primordial stochastic GW background, originating from quantum fluctuations in the early Universe, and amplified by an inflationary phase \cite{g76, g77, s80, l82}.  Standard inflationary models predict a primordial GW background whose amplitude is proportional to the energy scale of inflation~\cite{Mukhanov:1990me}.  Observations of primordial GWs therefore provide unique insights into poorly understood processes in the very early Universe and its evolution from $10^{-32}$~s after the Big Bang through to today.

In standard inflationary theories, the GW energy spectrum is expected to be nearly scale-invariant---above a certain frequency, the GW energy density decreases monotonically with increasing frequency~\cite{Mukhanov:1990me}.  The gravitational field has quantum mechanical fluctuations which are dynamic at wavelengths smaller than the cosmological horizon, $H^{-1} = \sqrt{3c^2/(8\pi G\rho)}$, and static due to causality at wavelengths larger than the horizon.  During inflation, modes are redshifted and pulled outside the horizon where their power is frozen in with an amplitude that corresponds to the size of the cosmological horizon, and hence, to the energy density of the Universe at that time.  As inflation progresses the energy scale of the Universe decreases, and the cosmological horizon grows.  This is a consequence of the null energy condition, which posits that the energy density of the Universe cannot increase as a function of time.  Modes that freeze out at larger physical wavelength have less power in them.  Therefore, the slowly and monotonically decreasing energy density of the Universe during inflation is responsible for the monotonically decreasing shape of the primordial power spectrum of all fields.  Spectra that decrease with increasing frequency are referred to as ``red'' spectra, and those that grow with increasing frequency are ``blue''.

A red spectrum, combined with observational constraints on the amplitude of GWs from the CMB, imply that GW detectors such as Pulsar Timing Arrays (PTAs) and ground-based interferometers such as the Laser Interferometer Gravitational-wave Observatory (LIGO) \cite{aligo2} and Virgo \cite{virgo} are not sufficiently sensitive to detect primordial GWs predicted by the simplest model of inflation \cite[e.g.,][]{abbott09}.  Detection at frequencies at or above PTAs may require extremely ambitious detectors such as the Big Bang Observer \cite{cutler06} or DECIGO \cite{seto01}.  However, some non-standard models for the early Universe predict blue GW spectra, which could be detected by PTAs and/or LIGO (see below).

A blue spectrum can be generated from inflation depending on what happens when GW modes exit the horizon, either by non-standard evolution of the Universe during inflation or if there is non-standard power in these modes when they exit.  This idea gained recent popularity in the wake of some early interpretations of the BICEP2 observations~\cite{Ade:2014xna}, where a flat GW spectrum was unable to simultaneously explain both the lower-frequency Planck observations~\cite{Ade:2015xua} and the higher-frequency BICEP2 results~\cite[e.g.,][]{mohanty14,mukohyama14,zhang14}.

Standard models of inflation suggest that the slope of the GW spectrum should be approximately equal to the slope of the power spectrum of density perturbations.  This prediction can be modified by having more than just a simple scalar field driving inflation.  These non-minimal models can predict either red GW spectra whose spectral index varies from that of standard inflation~\cite{Garriga:1999vw} or blue spectra \cite[e.g.,][]{piao04,baldi05}. The latter modification is so dramatic that the system violates the null-energy condition; a desirable, but by no means compulsory, property of the stress-energy tensor.  Alternatively, blue spectra can be generated if the propagation speed of primordial GWs varies during inflation \cite{cai15}, or by introducing new interactions between the scalar field and gravity, where these interactions are low-energy remnants of some (unknown) modification of general relativity at much higher energy scales such as the Planck scale.  Couplings of this form do not change any of the standard predictions of general relativity, but the theories that predict them allow us to treat the (unknown) high-energy theory of gravity in an {\it effective} low-energy limit for some energy scales.  The simplest of such effective field theories produce a blue spectrum~\cite{Baumann:2015xxa}.

It is also possible to abandon inflation altogether and replace it with a scenario which preserves the observed spectrum of density perturbations.  Two classic examples are string-gas \cite{brandenberger89} and ekpyrotic cosmologies \cite{Khoury:2001wf}.  In the former, an ensemble of fundamental strings have thermodynamic properties that produce a high-temperature, quasi-static state, which produces a blue GW spectrum, whose size is comparable in magnitude to the standard red spectrum~\cite{brandenberger14,brandenberger15}.  Ekpyrosis posits that the primordial spectrum of perturbations is a result of a pre-big bang contracting phase.  Such a phase has an increasing energy density and would create a blue power GW spectrum~\cite{Khoury:2001wf,Boyle:2003km}.

Importantly, a blue primordial GW spectrum may yield a primordial background immediately below present day limits, which may be detectable in the near future. While CMB experiments are likely to make direct measurements of the tensor-to-scalar ratio, $r$, they will poorly constrain the tensor index, $n_t$. In this paper, we show how the combination of constraints on the primordial GW background from CMB, PTA, BBN, BAO and ground-based interferometer GW experiments can place stringent constraints on $n_t$, yielding insights into the physics of the early Universe not accessible by any other means.

\section{Gravitational wave experiments}
Current results from experiments trying to measure the primordial GW background do little to constrain the possible tilt of the spectrum. However, combining GW experiments over all frequencies allows us to constrain cosmological parameters from non-standard inflationary cosmologies \cite{Boyle:2007zx,kuroyanagi15,Zhao:2013bba,meerburg15}. Combined CMB observations from the Planck satellite and the BICEP2 experiment constrain the stochastic GW background at frequencies of $\sim10^{-20}$--$10^{-16}$ Hz, while PTAs are sensitive to GW frequencies of  $\sim10^{-9}$--$10^{-7}$ Hz and ground-based interferometers are sensitive at $\sim10$--$10^3$ Hz  \cite{2000PhR...331..283M}.  As we show below, constraints on the total energy-density of GWs from BBN, gravitational lensing, CMB power spectra and BAO are sensitive to GWs as high as $10^9$~Hz.  Therefore, even a small blue tilt in the GW spectrum may be detectable in the GW frequency band covered from the CMB to LIGO/Virgo and provide more stringent constraints on the overall shape of the GW background~\cite{meerburg15}.

A first step in our effort to apply experimental constraints to the GW energy-density spectrum is to assume that it can be well-approximated by a power law:
\begin{align}
P_t(f) = A_t \left(\frac{f}{f_\cmb}\right)^{n_t}, \label{powerlaw}
\end{align}
where the pivot frequency, $f_\cmb$, is taken to be the standard value $f_\cmb = (c/2\pi)0.05\,{\rm Mpc}^{-1}$ \cite[e.g.,][]{cortes15}.  It is conventional to re-express the amplitude of the primordial GW spectrum in terms of the tensor-to-scalar ratio, $r\equiv A_t/A_s$, where $A_s$ is the amplitude of the primordial power spectrum of density perturbations, and both are evaluated at the pivot scale.  

Equation (\ref{powerlaw}) is the simplest approximation one can make about the primordial GW spectrum. 
Most early-Universe theories predict only a small deviation from pure power-law behavior.  The next level of complexity is to replace $n_t$ with $n_t+\alpha_t\ln(f/f_\cmb)/2$ in Eq. (\ref{powerlaw}), where $\alpha_t$ is known as the {\it running} of the spectral index.  For example, single-field, slow-roll inflationary models predict $\alpha_t \simeq (1-n_s)^2$ \cite[e.g., see][]{mukhanov92}, where $n_s =0.9645 \pm 0.0049$ is the measured value of the scalar spectral index~\cite{Ade:2015xua}. Therefore, within this class of theories, we expect a correction to the total GW power-law index of $\simeq10^{-2}$.  Over 29 decades in frequency, this can have a marginal effect on the results presented here, a point we discuss in more detail below.

Due to the expansion of the Universe, the primordial GW spectrum that we observe today has evolved since it was created.  This evolution is expressed in terms of a transfer function, $\mathcal{T}(f)$, which encodes information about how GWs change as a function of frequency \cite{Turner:1993vb}.  The energy density of GWs today is given by
\begin{align} 
  \rho_\gw = \int \frac{df f^4 (2\pi)^3}{c^5} P_t(f) \mathcal{T}(f)^2.
\end{align}

The PTA/LIGO communities commonly present the GW spectrum in terms of the energy density in GWs as a fraction of the closure energy density per logarithmic frequency interval~\cite{AllenRomano:1999, 2000PhR...331..283M}, 
\begin{equation}\label{eq:Omega}
  \Omega_\gw(f) \equiv \frac{1}{\rho_c} \frac{d \rho_\gw}{d\ln f} ,
\end{equation}
where $\rho_c \equiv 3 c^2 H_0^2/(8\pi G)$, $H_0 = 100 h$~km~$s^{-1}$~Mpc$^{-1}$ is the Hubble expansion rate, and $h=0.67$ is the dimensionless Hubble parameter~\cite{Ade:2015xua}.  Indirect constraints on the GW background are typically ``integral bounds'' on $\Omega_\gw\equiv\int d\ln f\Omega_\gw(f)$.

Assuming a standard expansion history that includes non-relativistic matter and radiation, the GW spectrum today is given by~\cite{Turner:1993vb,Watanabe:2006qe,Smith:2008pf,Smithinprep}:
\begin{align}
	\Omega_\gw(f)=\Omega_\gw^\cmb\left(\frac{f}{f_\cmb}\right)^{n_t}\left[\frac{1}{2}\left(\frac{f_\eq}{f}\right)^2+\frac{16}{9}\right],\label{omega_gw}
\end{align}
where $f_\eq$ is the frequency of the mode whose corresponding wavelength is equal to the size of the Universe at the time of matter--radiation equality with frequency $f_\eq=\sqrt{2}cH_0\Omega_m/2\pi\sqrt{\Omega_r}$.  Here, $\Omega_m$ and $\Omega_r$  are respectively the total matter and radiation energy-density evaluated today, and
\begin{align}
	\Omega_\gw^\cmb\equiv 3rA_s\Omega_r/128.\label{omega_gw_r}
\end{align}
Equation (\ref{omega_gw}) is key in our analysis since it allows us to combine constraints on $r$ and $n_t$ from the CMB with constraints to $\Omega_\gw(f)$ and $n_t$ from PTAs and LIGO.  We use cosmological parameters obtained from the latest Planck satellite data release~\cite{Ade:2015xua}.

\begin{figure*}
\includegraphics[width=1.95\columnwidth]{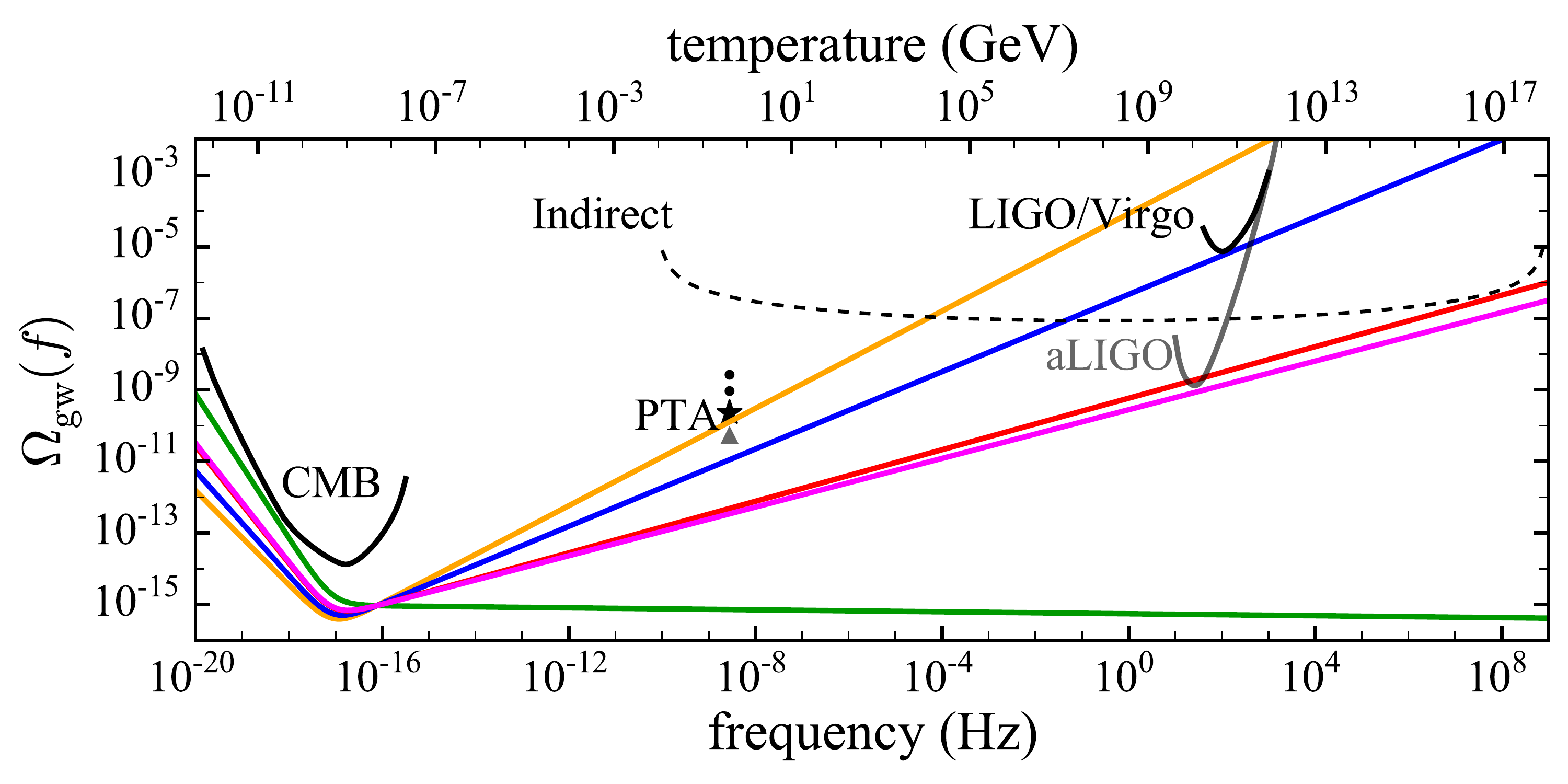}
\caption{\label{Omega_freq} Experimental constraints on $\Omega_{\rm gw}(f)$;  the black star is the current PPTA upper limit and all black curves and data points are current 95\% confidence upper limits. The grey curve and triangle are respectively the predicted aLIGO sensitivity and PPTA sensitivity with five more years of data. The indirect GW limits are from CMB temperature and polarization power spectra, lensing, BAOs, and BBN.
    Models predicting a power-law spectrum that intersect with an observational constraint are ruled out at $>95\%$ confidence. 
   We show five predictions for the GW background, each with $r=0.11$, and with $n_t=0.68$ (orange curve), $n_t=0.54$ (blue), $n_t=0.36$ (red), $n_t=0.34$ (magenta), and the consistency relation, $n_t=-r/8$ (green), corresponding to minimal inflation.   
}  
\end{figure*}

In the following sections, we combine observational constraints spanning 29 orders of magnitude in frequency to derive stringent constraints on backgrounds with a non-zero spectral index $n_t$.  Figure \ref{Omega_freq} highlights the key idea: we show the current best upper limits on $\Omega_\gw(f)$, and a series of curves given by Eq.~(\ref{omega_gw}) that are constrained by these limits.  Starting from the lowest frequency limits, we summarize current upper limits before combining them to derive joint constraints.

Note that in evaluating the observed GW spectrum we have neglected the effects of neutrino free-streaming \cite{Weinberg:2003ur} and phase transitions occurring in the early universe \cite{Watanabe:2006qe}.  As shown in Refs. ~\cite{Weinberg:2003ur,Watanabe:2006qe,liu15}, the free-streaming of neutrinos and phase transitions during the very early Universe lead to a suppression of $\Omega_{\rm gw}$ by a factor of $\sim 1/2$ -- $1/3$ for PTA and LIGO frequencies.  However, because of the large lever-arm between the frequencies probed by these detection methods and the CMB, including this suppression in the analysis changes our constraints on $n_t$ by only a few percent.  We therefore neglect these effects in our analysis. 

Three recent papers have presented combined constraints on $n_t$ and $r$ using some combination of CMB, LIGO and PTA data \cite{meerburg15, huang15a, liu15}\footnote{After our paper was submitted to the journal, \citet{cabass15} also presented combined constraints on $n_t$ and $r$, as well as forecasts on the capability of future CMB satellite experiments to constrain $n_t$.}.  \citet{huang15a} presented their analysis soon after the original BICEP2 results were reported~\cite{aab+14,ade15}, and as such, focussed on the fact that those data preferred a slightly positive blue tilt for the tensor power spectrum, which resulted from the inconsistency between the original BICEP2 data and Planck observations.  On the other hand, \citet{liu15} presented constraints from only CMB and PTA data, but focussed on what a positive detection could do for our understanding of the early-Universe equation of state, cosmic phase transitions and relativistic free-streaming.

Our analysis improves on those of Refs.~\cite{meerburg15,huang15a,liu15} in a number of significant ways.  First, we include the indirect GW constraints in a {\it self-consistent} way, which allows us to compare integral and non-integral constraints with varying spectral indices (see Section \ref{indirect}).  Second, we present a new analysis of PPTA data \cite{shannon15} that give the best limit on $\Omega_\gw(f)$ in the PTA band by a factor of four over previous published results.  Finally, we provide our own analysis of the raw PPTA time-of-arrival data to allow for varying spectral indices, instead of assuming a constant $n_t$ for PTA observations taken from older PPTA analyses as is done in~\cite{meerburg15, liu15}.

\subsection{CMB Intensity and Polarization}
Primordial GWs imprint a characteristic signal onto the intensity and polarization of the CMB that can be measured by ground-based and space-borne observatories.  A joint analysis~\cite{Ade:2015xua,ade15} of Planck satellite and BICEP2/Keck array data found that $r< 0.12$ at 95\% confidence level (CL) under the assumption that $n_t = 0$.  The solid black curve in Fig.~\ref{Omega_freq} labeled ``CMB''  shows the estimated sensitivity of the Planck satellite.  The CMB sensitivity curve is calculated by determining the value of the spectral density $\Omega_{\gw}(f)$ that yields a marginally detectable signal given a model of the Planck satellite noise properties \cite{2011A&A...536A...1P}.  

Observations of the CMB intensity and polarization are analyzed by re-expressing the real-space data in a spherical harmonic expansion.  The intensity measurements can be expanded in the standard (scalar) spherical harmonics, whereas the polarization data must be expanded in spin-weighted spherical harmonics \cite{Kamionkowski:1996ks}.  We can further separate various physical processes by dividing the polarization data into a curl-free (E-mode) and curl (B-mode) basis. In order to compare these data to a theoretical model, the measured spherical harmonic coefficients are further analyzed to estimate their statistical correlations.  The presence of a primordial GW spectrum fundamentally alters the expected correlations leading to an enhanced correlation for the intensity of the CMB on the largest angular scales as well as a non-zero correlation for the B-mode polarization \cite{Kamionkowski:1996zd,Seljak:1996gy}.  

The expected effect of a non-zero primordial GW spectrum on the CMB is calculated by solving the Boltzmann equation for the various components of matter that fill the Universe.  The Boltzmann equation for the photons encodes all of the information about correlations in the intensity and polarization of the CMB.  In particular, for each spherical harmonic multipole, the expected correlations can be expressed as an integral over cosmic time and frequency \cite{Seljak:1996is}.  Therefore, the total expected CMB signal due to primordial GWs can be expressed as an integral over its spectrum, $\Omega_{\gw}(f)$.  

The CMB sensitivity curve shown in Fig.~\ref{Omega_freq} is calculated by setting the total CMB signal-to-noise ratio equal to two (corresponding to a 95\% CL bound).  The squared signal-to-noise ratio is calculated from a sum in quadrature of CMB B-mode multipoles divided by the estimated polarization noise for each multipole from the Planck satellite's 143 GHz detector \cite{2011A&A...536A...1P,planckbluebook}.  Since, as discussed above, each B-mode multipole is an integral over the GW spectrum, we express the integral over frequency as a sum so that we can evaluate the contribution at each frequency interval. We relate the primordial amplitude to the present-day spectral density using Eq.~(21) of~\cite{Watanabe:2006qe}. The noise is calculated under the hypothesis of no primordial GWs, although weak gravitational lensing also induces a non-zero B-mode correlation, which we treat as an additional source of noise.  Finally, the limit is converted into a ``power-law integrated'' curve using the formalism from~\citet{thrane13}.  Any model intersecting this curve is ruled out at 95\% CL.

Current CMB constraints to $r$ and $n_t$ come from measurements made by Planck~\cite{Aghanim:2015xee}, the BICEP2/Keck array \cite{ade15}, and SPTpol \cite{Keisler:2015hfa}.  Constraints from these datasets were determined using the Boltzmann solver CAMB and a modified version of the Monte Carlo stepper cosmoMC \cite{Lewis:1999bs,Lewis:2002ah,Lewis:2013hha}.

\subsection{Pulsar Timing Arrays}
\label{PPTA}
The incoherent superposition of primordial GWs is expected to imprint on the arrival time of pulses from the most stable millisecond pulsars.  A number of PTAs around the world are engaged in the hunt for GWs, including the Parkes Pulsar Timing Array \cite[PPTA; ][]{manchester13}, the North American Nanohertz Observatory for Gravitational Waves \cite[NANOGrav;][]{mclaughlin13}, and the European Pulsar Timing Array \cite[EPTA; ][]{kramer13}.  Here, we use recent data from the PPTA \cite{shannon15} to provide the strongest constraints to date on $\Omega_\text{gw}(f)$ from a primordial background in the PTA band.

The PPTA monitors 24 pulsars with the 64-m Parkes radio telescope in a bid to directly detect GWs, and currently has the most stringent upper limits on the GW background from supermassive black hole binaries~\cite{shannon15}.  We derive our limit on the primordial GW background by performing a similar Bayesian analysis to that in~\cite{shannon15}, with the exception that we utilize the Bayesian pulsar timing data analysis suite PAL2\footnote{\href{https://github.com/jellis18/PAL2}{https://github.com/jellis18/PAL2}.  We note that PAL2 gives consistent results to the analysis in \citet{shannon15}.}, and allow for an arbitrary strain spectral index.

The GW spectrum in the PTA band can be approximated as a power law, with  
\begin{align}
	\Omega_\gw(f)=\frac{2\pi^2}{3H^2_0}{A_\gw^2f_\yr^2}\left(\frac{f}{f_\yr}\right)^{n_t},\label{Omega_PTA}
\end{align}
where $A_\gw$ is the amplitude of the characteristic strain at a reference frequency of $f_\yr\equiv{\rm yr}^{-1}$.
The star in Fig.~\ref{Omega_freq}, labelled ``PTA'', is the 95\% CL upper limit assuming a spectral index of $n_t=0.5$ (approximately the middle of the range we are trying to constrain---see below), with $\Omega_\gw^{95\%}(f)<2.3\times10^{-10}$.
The black dots above the PPTA limit are the upper limits  from the EPTA \cite{lentati15} and NANOGrav \cite{arzoumanian15}.  Both the EPTA and NANOGrav present limits on the GW energy density from inflationary relics assuming $n_t=0$; our new limit for $n_t=0$ (cf. our limit on $\Omega_\gw(f)$ for $n_t=0.5$ which only differs in the second decimal place) is a factor of 4.1 better than the previous best limit from~\cite{arzoumanian15}. 

The grey triangle below the star in Fig.~\ref{Omega_freq} is a predicted GW upper limit derived by simulating an additional five years of PPTA data.  We took the maximum likelihood red noise parameters in the existing data sets, estimated the white noise level using the most recent data that represents current observation quality, and assumed a two-week observing cadence to derive the 95\% CL upper limit of $\Omega_\gw^{95\%}(f)\lesssim 5\times10^{-11}$. However, the PPTA limit will be superseded before 2020 with limits placed from collating datasets from the three existing PTAs as part of the International Pulsar Timing Array \cite[IPTA; ][]{2010CQGra..27h4013H}.  From Fig.~\ref{Omega_freq}, it becomes clear that PTAs may not play a significant role in constraining inflationary models where the GW spectrum is described by Eq. (\ref{eq:Omega}) when aLIGO reaches design sensitivity, given the significant improvements in the latter experiment. However, PTAs can still play an important role for cosmological models with a varying spectral index; that is, with a non-negligible running of the spectral index $\alpha_t$.

 \citet{giblin14} recently proposed a ``rule of thumb'' for the maximum GW energy density for cosmological backgrounds based only on arguments of the energy budget of the Universe at early times.  They presented optimistic, realistic and pessimistic upper limits for $\Omega_\gw(f)$, with the optimistic limit representing the largest value of $\Omega_\gw(f)$ possible given a reasonable set of conditions.  The new PPTA limit reported here is the first time a GW limit in either the PTA or LIGO band has gone under this optimistic threshold, thus marking the first time the detection of cosmological GWs could actually have been possible according to arguments in \cite{giblin14}.  
 Conventional models of early-Universe particle physics do not predict such a large GW background in the PTA frequency band.
The temperature of the Universe at the time when such GWs are produced is $\sim1\,{\rm GeV}$ (see top axis of Fig.~\ref{Omega_freq}), a temperature at which physics of the early Universe is relatively well known.  
We note that the possibility of first-order phase transitions that generate a strong GW background in the PTA frequency band is not completely ruled out~\cite[e.g.,][]{caprini10,giblin14b, kalaydzhyan15, child15}  Of course it is possible that there is {\it unknown} physics that influences gravity without coupling strongly to the standard model of particle physics that could produce a strong GW background in the PTA frequency band.

\begin{figure*}[ht]
\includegraphics[width=2.0\columnwidth]{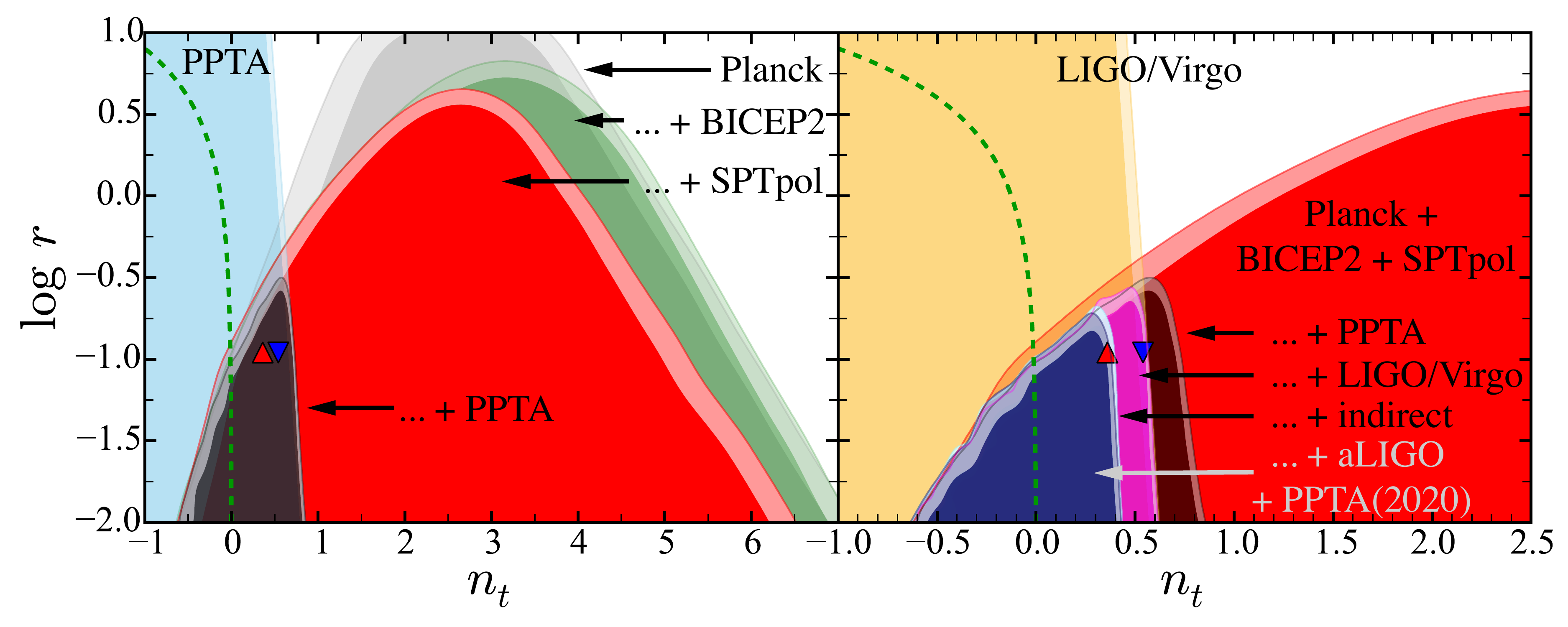}
	\caption{\label{combined_posterior} Combined, two-dimensional posterior distribution for the tensor-to-scalar ratio, $r$, and the blue tilt of the GW spectrum, $n_t$, using CMB, PPTA, indirect, and LIGO observations.  The contours are the $95$ and $99\%$ limits.  The green, dashed curve shows the consistency relation, $n_t=-r/8$, while the red and blue triangles correspond respectively to the red and blue curves in Fig \ref{Omega_freq}.  For clarity, the right panel is a zoomed-in version of the left panel, with additional posterior distributions shown.  See Section \ref{results} for a description of each posterior distribution. 
} 
\end{figure*}

\subsection{Ground-based interferometers}
\label{LIGO}
LIGO~\cite{aligo2} and Virgo~\cite{virgo} are long-baseline, ground-based GW interferometers with best sensitivity at frequencies of $10^2$--$10^3$ Hz.  Data collected during the initial phases of these instruments have been used to place upper limits on a stochastic background of GWs from astrophysical and cosmological sources \cite{abbott09,aasi14,aasi15}.  We utilize data from the initial LIGO and Virgo observatories.  These data were collected in 2009 -- 2010 as part of the fifth LIGO Science run.  Two limits were originally obtained using these combined observations: a lower-frequency limit from combined LIGO-Virgo observations that assumed a flat, i.e., $n_t=0$, spectrum~\cite{aasi14}, and a higher-frequency limit from an analysis of the two co-located LIGO detectors at Hanford, which assumed $n_t=3$ \cite{aasi15}.

We implement a new way to analyze LIGO/Virgo limits on the primordial background that allows for a varying spectral index.  The analysis goes beyond \citet{meerburg15}, and \citet{huang15a}, which both assume $n_t=0$ for their LIGO/Virgo constraints.  We combine published data from Refs. \cite{aasi14} and \cite{aasi15} to generate a power-law integrated curve \cite{thrane13}, shown in Fig.~\ref{Omega_freq}.  Any power-law model intersecting a power-law integrated curve is ruled out at 95\% CL.  Then, utilising the formalism from~\citet{mandic12}, we obtain constraints on $\Omega_\text{gw}(f)$ for arbitrary spectral indices.  The limits on $\Omega_\text{gw}(f)$ are converted into constraints on $n_t$ and $r$.

At the time of writing, the LIGO experiment has begun taking data for the first observing run of the advanced detector era, with the Virgo experiment to follow in 2016.  At design sensitivity, advanced detectors are forecast to achieve nearly four orders of magnitude of improvement in $\Omega_\text{gw}(f)$; see the curve marked ``aLIGO'' in Fig.~\ref{Omega_freq}, which is the projected sensitivity given two LIGO detectors operating for one year at design sensitivity. 

\subsection{Indirect constraints}
\label{indirect}
Indirect constraints on GW backgrounds have been obtained using a variety of data including CMB temperature and polarization power spectra, lensing, BAOs, and BBN \cite[e.g.,][]{Maggiore:1999vm,Smith:2006nka,Sendra:2012wh}. Indirect bounds are ``integral bounds,'' which apply to $\Omega_\text{gw}$ and not to $\Omega_\text{gw}(f)$; see Eq.~\ref{eq:Omega}.  Recently, \citet{Pagano:2015hma} combined the latest Planck observations of CMB temperature and polarization power spectra and lensing with BAO and BBN measurements (specifically observations of the primordial Deuterium abundance) to put an integral constraint on the primordial GW background of $\Omega_\gw<3.8\times10^{-6}$.  

While there is a long history in the literature of plotting $\Omega_\gw$ integral bounds alongside $\Omega_\text{gw}(f)$, they are not directly comparable.
However, the two quantities can be related if we assume that $\Omega_\text{gw}(f)$ is described by a power-law spectrum with a known cut-off frequency, which we choose to be $f_{\rm max}=1$ GHz, corresponding to an energy scale typical of inflation, $T=10^{17}$~GeV.
Given this plausible assumption, we plot the indirect constraints in Fig.~\ref{Omega_freq} as power-law integrated curves using the formalism from Ref.~\cite{thrane13}.
Any power-law model intersecting a power-law integrated curve is ruled out at 95\% CL.

Inspecting Fig.~\ref{Omega_freq}, it is apparent that the current best constraints on $n_t$ come from observations of the CMB combined with indirect bounds.  

The strength of the indirect bounds depends in part on our choice of $f_{\rm max}=1$~GHz, however changing the cut-off frequency by several orders of magnitude would not the change qualitative picture.  For example, in alternative theories of inflation it is possible to posit an energy scale as low as $10^6$ GeV, corresponding to a cut-off frequency of $f_{\rm max}=10$ mHz.  This choice of cut-off frequency shifts the minimum of the indirect bound curve from $\sim 100$ mHz to $\sim1\,\mu$Hz, while the minimum value of $\Omega_{\rm gw}(f)$ increases by a factor of $\sim2$.  When aLIGO reaches design sensitivity, it will surpass indirect constraints on primordial backgrounds with non-running spectral indices.

\section{Combined constraints on the primordial tilt}
\subsection{Combined experimental constraints}
\label{results}

Here we combine the current limits on $\Omega_\text{gw}(f)$ from the individual experiments mentioned above to constrain the tensor-to-scalar ratio, $r$, and the tensor index, $n_t$.  In Fig.~\ref{combined_posterior}, we plot these two-dimensional posterior distributions for $r$ and $n_t$.  In both panels we plot two theory points and a theory curve.  The green, dashed curve corresponds to the consistency relation from standard inflationary models ($n_t=-r/8$), and the red and blue triangles have the same values of $n_t$ and $r$ as the red and blue curves in Fig.~\ref{Omega_freq}.

Figure \ref{combined_posterior} combines the constraints from each experiment in a heuristic manner.  The left panel shows all of the CMB constraints from direct detection experiments starting with Planck (grey shaded region), adding BICEP2 (green), and finally adding SPTpol (red) to get the overall constraints on the CMB from direct GW observations.  Also plotted in the left panel is the PPTA posterior (blue).  The PPTA search algorithm described in Section \ref{PPTA} derives a posterior distribution in terms of $n_t$ and $\Omega_\gw(f)$, which is converted to $n_t$ and $r$ using Eqs.~(\ref{omega_gw}) and (\ref{omega_gw_r}).  Finally, in the left-hand panel of Fig.~\ref{combined_posterior}, we plot the combined CMB + PPTA posterior (black).  This distribution represents the state-of-the-art constraints one can derive from CMB and PTA experiments alone.  At a reference value of $r=0.11$, this limits $n_t<0.68$ with 95\% confidence.  In general, the 95\% CL upper limit on $n_t$ as a function of $r$ derived from these constraints is well-approximated by the simple relation
\begin{align}
	n_t=A\log_{10}\left(\frac{r}{0.11}\right)+B,\label{ntr_eqn}
\end{align}
where $A=-0.13$ and $B=0.68$.  Equation (\ref{ntr_eqn}) allows one to extrapolate the constraints on $n_t$ to arbitrary small values of $r$.  With the addition of each experimental constraint we get tighter limits on $A$ and $B$; the 95\% CL upper limits for each experiment are collated in Table \ref{table}.

In the right panel of Fig.~\ref{combined_posterior} we retain the CMB (red, labelled `Planck + BICEP2 + SPTpol') and CMB + PPTA (black) posterior distributions from the left-hand panel.  As with the PPTA analysis, the LIGO/Virgo data analysis algorithm described in Section \ref{LIGO} constrains $n_t$ and $\Omega_\gw$, which we convert to $n_t$ and $r$ using Eqs.~(\ref{omega_gw}) and (\ref{omega_gw_r}).  This distribution is shown in yellow, and the combined CMB + PPTA + LIGO/Virgo constraints are shown in pink.  This pink contour represents the current best constraint from the direct GW experiments covering 21 decades in frequency.  At a reference value of $r=0.11$, the CMB + PPTA + LIGO/Virgo constraints yields an upper limit of $n_t<0.54$.  For smaller values of $r$, the theoretical curves at the CMB frequency take on lower values of $\Omega_\gw$, which implies higher-frequency experiments play even more of a role in constraining $n_t$ than comparatively lower-frequency experiments.  For example, at $r=0.01$, CMB + PPTA constraints imply $n_t<0.82$, while CMB + LIGO/Virgo constraints imply $n_t<$0.60.  The 95\% CL from the CMB + LIGO constraint is well approximated by Eq. (\ref{ntr_eqn}) where values for $A$ and $B$ can be found in Table \ref{table}.

Also in the right panel of Fig. \ref{combined_posterior}, we add the indirect constraints described in Section \ref{indirect} to the direct constraints (turquoise).  This contour represents our total knowledge of $n_t$ and $r$ using all experimental constraints.  In this case, at $r=0.11$ we find $n_t<0.36$ at 95\% CL, and the upper limit as a function of $r$ is well approximated by Eq. (\ref{ntr_eqn}), where the values for $A$ and $B$ can be found in Table \ref{table}.

From Fig.~\ref{combined_posterior}, it is clear that only direct observations of the CMB constrain $n_t<0$.  This makes sense in the context of Fig. \ref{Omega_freq}: given that the lever-arm for the GW theory curves, i.e., Eq. (\ref{omega_gw}), are hinged at $f_\cmb$, a negative spectral index is not constrained by experiments that are only sensitive to values of $\Omega_\gw(f)$ higher than at the CMB.

Finally, in Fig. \ref{combined_posterior} we show the projected constraints that one can expect by the year 2020 (dark blue, labelled ``\ldots + aLIGO + PPTA(2020)'') assuming five more years of PPTA observations and aLIGO at design sensitivity (see Sections \ref{PPTA} and \ref{LIGO}).  These contours show that the constraint for the spectral index improves to $n_t\lesssim0.34$ at $r=0.11$, and is well-approximated by Eq. (\ref{ntr_eqn}) with $A$ and $B$ given in Table \ref{table}.  As is evident from Fig. \ref{Omega_freq}, the constraint at high $n_t$ will be dominated by aLIGO.  Similar constraints in the PTA band are not expected until the era of the Square Kilometre Array and Five hundred meter Aperture Spherical Telescope (FAST)~\cite[e.g.,][]{janssen+14, liu15}.

\begin{center}
\begin{table}[b]
	\begin{tabular}{|c|cc|}
	\hline
	Experiment & A & B \\
	\hline
	CMB + PPTA & -0.13 & 0.68\\
	CMB + PPTA + LIGO & -0.06 & 0.54\\
	CMB + PPTA + LIGO + indirect & -0.04 & 0.36\\
	CMB + PPTA(2020) + aLIGO & -0.06 & 0.34	\\
	\hline
	\end{tabular}
	\caption{\label{table}95\% CL upper limits on $A$ and $B$ as in Eq. (\ref{ntr_eqn}).  The value of $B$ is therefore the 95\% upper limit of $n_t$ at a reference value of $r=0.11$.}
\end{table}
\end{center}

\subsection{Comparison with theory}
\label{theory_compare}
In the previous subsection, we present stringent constraints on the blue tilt of the primordial GW background from experiments spanning 29 decades in frequency.  These results can be used to comment on early Universe models.  Those models whose spectral indices are near zero -- or of comparable magnitude to standard inflationary models -- are consistent with the data.  String-gas cosmologies and modified inflationary scenarios with non-minimal couplings to gravity seem to be the least constrained, since these models predict relatively small values of $n_t$ and are unconstrained for even relatively large tensor-to-scalar ratios.  

Ekpyrosis has a tendency to predict large values of the spectral tilt including $n_t\approx2$ \cite{Khoury:2001wf} that come from modes freezing out of the horizon during the contracting phase of the Universe.  More modern incarnations of ekpyrosis produce blue tilts, but with relatively low values of tensor-to-scalar ratio, $r$, \cite[e.g.,][]{Boyle:2003km}.  Here, our derivation of fitting formulae for $n_t$ as a function of $r$ (see Table \ref{table}) allow specific ekpyrotic predictions to be tested to arbitrary small values of $r$.

Our results will have important implications following the detection of non-zero tensor-to-scalar ratio by a future CMB experiment (e.g., see Ref.~\cite{huang15b}\footnote{One should be cautious of the projected constraints from \citet{huang15b}.  Their Fisher matrix analysis necessarily assumes the posterior distribution is Gaussian, and hence symmetric about some fiducial model.  This is not the case for $n_t$, which allows for significantly larger positive values than it does negative.} and references therein).  Such a detection, together with the data from PTAs and ground-based interferometers will put very tight limits on $n_t$, with larger values of $r$ being the most constraining.  For example, a confirmed detection of $r\approx 0.1$ would put very tight bounds on $n_t$ with a strong preference for {\it small} and {\it positive} values.  Such tight constraints are truly a result of CMB bounds on the low frequency end and PTA, LIGO and indirect bounds on the upper end.

When a detection is made in any of the frequency bands studied herein, it becomes even more pertinent to analyze all experimental data, consistently taking into account the spectral running, $\alpha_t$.  Indeed, in the case of a detection, upper limits in each frequency band can be used to simultaneously constrain $n_t$ and $\alpha_t$; three or more experiments are required to constrain both parameters.  In a future work, we will present three-dimensional posterior constraints that include $r$, $n_t$ and $\alpha_t$, and also incorporate predictions for future CMB experiments.  Distinguishing primordial backgrounds from astrophysical foregrounds may be a daunting task, though multi-wavelength measurements could prove useful toward this end.

\section{Conclusion}
\label{conclusion}
By combining limits from many different GW experiments probing 29 decades in frequency, we present new constraints on cosmological parameters $n_t$ and $r$, which are intimately related to the evolution of the early Universe.  This interdisciplinary research also makes significant advances in PTA, LIGO/VIRGO and indirect GW limit analysis techniques. Specifically, we present new PPTA data that provides the most stringent limit on the primordial gravitational-wave background, $\Omega_\gw(f) <2.3\times10^{-10}$; more than a factor of four tighter than the previous best limit from~\cite{arzoumanian15}.  Moreover, we develop and implement a method to give the best limits on the primordial background from ground-based interferometers---a method we anticipate will become standard in future LIGO/Virgo primordial background analyses. Furthermore, we provide a new interpretation of indirect GW constraints from CMB temperature and polarization measurements, lensing, BBN and BAO observations that allow for a varying primordial spectral index, allowing us to directly compare these ``integral'' constraints on $\Omega_\gw$ with the usual frequency-dependent $\Omega_\gw(f)$ constraints.  Our technique for comparing direct and indirect limits can be widely adopted within the GW community to avoid the confusion created from `apples-to-oranges' comparisons.

While Refs. \cite{meerburg15, huang15a, liu15} present constraints on $n_t$ and $r$ using combinations of CMB, LIGO and PTA data, the focus of their work was significantly different.  Indeed, the work of \citet{meerburg15} and \citet{huang15a} were originally in response to the now defunct BICEP2 results~\cite{aab+14,ade15}, while \citet{liu15} presented constraints from only CMB and PTA data, but focussed on what a positive detection could do for our understanding of the early-Universe equation of state, cosmic phase transitions and relativistic free-streaming. 

A direct comparison between our results and that of \citet{meerburg15} is not possible for a number of reasons.  Notably, they use a linear prior on $r$, which, together with the use of the original BICEP2 results, ends in constraints that are not bounded below.  Figure~\ref{Omega_freq}, together with Eqs.~(\ref{omega_gw}) and (\ref{omega_gw_r}), show that $r$ should be unbounded below given that {\it there are no lower limits} on the amplitude of $\Omega_\gw(f)$ from any experiments.  Moreover, \citet{meerburg15} use an unconventional pivot scale for the theoretical GW spectrum (see, e.g., Ref.~\cite{cortes15} for a discussion of the optimal pivot scale).  

Our results are significantly more constraining than those of \citet{liu15} (see their Fig.~7), most notably due to the inclusion of indirect GW constraints.  Our analysis quantifies how large the spectral index of the primordial spectrum, $n_t$, can grow as a function of the tensor-to-scalar ratio, $r$; see Fig. \ref{combined_posterior} and Table \ref{table} for a summary of the results. 

Various theories of the early Universe predict a blue primordial gravitational-wave spectrum \cite[e.g.,][]{Khoury:2001wf,Boyle:2003km,baldi05,Baumann:2015xxa}, and indeed, some versions of ekpyrosis predict large values of $n_t$ which we can now rule out by our analysis -- see Section \ref{theory_compare}.  Observations of the CMB alone only limit the inflationary GW spectrum to $n_t\lesssim5$ at a reference value of the tensor-to-scalar ratio of $r=0.11$.  Current observations by the PPTA and initial LIGO/Virgo reduce this limit to $n_t<0.54$ with 95\% confidence, and including limits from indirect GW observations reduces this to $n_t<0.36$.  We predict that observations by aLIGO at design sensitivity (circa 2020) will reduce this constraint to $n_t<0.34$.  All upper limits on $n_t$ are applicable at a reference value of $r=0.11$, but can be extrapolated to other values of $r$ using Eqn.~(\ref{ntr_eqn}) and Table~\ref{table}.

Of course, it is a future {\it direct detection} of $r$ that will have the most important implications.  Such a detection will allow us to slice through the parameter space presented in Fig.~\ref{combined_posterior}, providing significant constraints on parameters governing theories of the early Universe.

\acknowledgments
PDL is grateful to Justin Ellis for valuable support with the pulsar timing package PAL2.  
We are extremely grateful to the three referees who all provided excellent feedback that improved the manuscript.
This work was initiated at the Aspen Center for Physics, which is supported by National Science Foundation grant PHY-1066293.
JTG is supported by the National Science Foundation, PHY-1414479.
CMFM was supported by a Marie Curie International Outgoing Fellowship within the European Union Seventh Framework Programme. RRC is supported in part by DOE grant DE-SC0010386.
The Parkes radio telescope is part of the Australia Telescope National Facility which is funded by the Commonwealth of Australia for operation as a National Facility managed by CSIRO.
YL and GH are recipients of ARC Future Fellowships (respectively, FT110100384 and FT120100595). YL, MB, WvS, PAR and PDL are supported by ARC Discovery Project DP140102578. SO is supported by the Alexander von Humboldt Foundation. LW and XZ acknowledges support from the ARC.  J-BW is supported by NSFC project No.11403086 and West Light Foundation of CAS XBBS201322. XPY is supported by NSFC project U1231120, FRFCU project XDJK2015B012 and China Scholarship Council (CSC).

\bibliography{references}

\end{document}